\def\funp{{I\!\!P}}
\def\xp{x_{{I\!\!P}}}

\newcommand{\be}{\begin{equation}}
\newcommand{\ee}{\end{equation}}
\newcommand{\beeq}{\begin{eqnarray}}
\newcommand{\eeeq}{\end{eqnarray}}

\documentclass{appolb}
\usepackage{epsfig}

\begin{document}
\title{Saturation and Diffractive DIS%
\thanks{Presented at the XXXIII International Symposium
on Multiparticle Dynamics,
Krak\'ow, Poland, 5-11 September 2003.
 }%
}
\author{K. Golec-Biernat
\address{H. Niewodnicza\'nski Institute of Nuclear Physics, Cracow, Poland}
}
\maketitle
\begin{abstract}
We review QCD based descriptions of diffractive deep inelastic scattering
emphasizing the role of models with  parton saturation.
These models provide natural explanation of such experimentally observed facts as the constant
ratio of $\sigma^{diff}/\sigma^{tot}$ as a function of the Bjorken variable $x$ and Regge
factorization of diffractive parton distributions.
\end{abstract}

\section{Introduction}

Around $10\%$ of deep
inelastic scattering (DIS) events observed at HERA at small value of the Bjorken variable $x$ are
diffractive events \cite{H197,ZEUS99}, when the incoming proton
stays intact  losing only a small fraction $\xp$ of its initial momentum.
A large rapidity gap is formed
between the scattered proton (or its low mass excitation) and the diffractive system.
The ratio of diffractive to total DIS cross sections
is to a good approximation constant as a function of $Q^2$.
Thus in a first approximation, DIS diffraction is a leading
twist effect with logarithmic scaling violation. Moreover, the same ratio as a function
of $x$ (or energy) is also constant. Theoretical models of diffraction
should explain these facts.

\section{Diffractive parton distributions}

\begin{figure}[t]
  \vspace*{-0.5cm}
     \centerline{
         \epsfig{figure=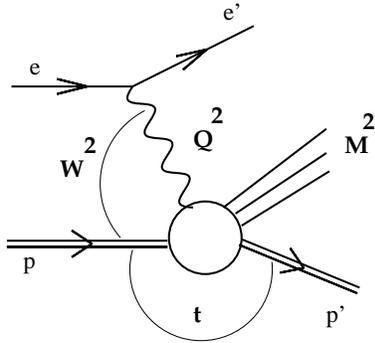,width=5cm}
           }
\vspace*{0.0cm}
\caption{\it Kinematic invariants in DIS diffraction.
\label{fig:d1}}
\end{figure}

In addition to the Bjorken variable $x=Q^2/(Q^2+W^2)$, there are two dimensionless
variables used in the description of DIS diffraction
\be
\label{eq:1}
x_\funp=\frac{Q^2+M^2}{Q^2+W^2}\,,~~~~~~~~~~~~~\beta=\frac{x}{x_\funp}=\frac{Q^2}{Q^2+M^2}\,,
\ee
where $M^2$ is invariant mass squared of the diffractive system
and $W^2$ is the center-of-mass energy squared of the $\gamma^*p$ system, see Fig.~\ref{fig:d1}.
In analogy to the inclusive
DIS, the diffractive structure functions are defined: $F^D_{2,L}(x,\xp,Q^2,t)$, where
$t=(p-p^{\prime})^2$ is the four momentum squared transferred from the proton  into the diffractive system.

The leading twist description of diffractive DIS is realized using the diffractive
parton distributions (DPD) $q^D$ in terms of which
\be
\label{eq:2}
F^D_{2}=\sum_{i=1}^{N_f}e_i^2\,\beta\,\left\{q^D_i(\xp,t;\beta,Q^2)+
\overline{q}^D_i(\xp,t;\beta,Q^2)\right\}\,,
\ee
where $i$ enumerates quark flavours. Eq.~(\ref{eq:2}) is an example of the collinear factorization formula
proven for DIS diffraction in \cite{COL98}. In the infinite momentum frame, the DPD have an interpretation
of conditional probabilities to find a parton in the proton
with the momentum fraction $x=\beta \xp$ under the condition that the incoming proton stays intact and loses
the fraction $\xp$ of its  momentum.
The collinear factorization fails in hadron--hadron hard diffractive scattering
due  to initial  state soft interactions \cite{CFS93,REVWU}. Thus, unlike
inclusive scattering,  the diffractive parton distributions are no universal
quantities. They can be used, however,  for different diffractive processes in lepton--nucleon
scattering, e.g. for diffractive dijet production.

The collinear factorization theorem
of \cite{COL98} allows to use the Altarelli-Parisi (DGLAP) evolution equations to find the $Q^2$
dependence of DPD, provided the initial conditions for evolution are known. They are found from
fits to diffractive DIS data in full analogy to  the determination of
inclusive parton distributions \cite{H197,ZEUS99}.
In the evolution equations only $(\beta, Q^2)$ are relevant variables while
$(\xp,t)$ play the role of external parameters. Thus a modelling of the latter dependence for
the DPD is necessary. This is done using physical ideas about the nature of interactions
leading to DIS diffraction.

Traditionally, diffraction is related to  the exchange of a pomeron:   a vacuum quantum number
exchange, described by the linear Regge trajectory
$\alpha_\funp(t)=\alpha_\funp(0)+\alpha^\prime t$ with $\alpha_\funp(0) \ge 1$, which dominates
at high energy ($s\to \infty,\, t=\mbox{\rm const}$). This is the basis of the Ingelman--Schlein (IS)
\cite{IS} model in which the  pomeron is exchanged between the proton and the diffractive system. In this case
$F^D_2$ factorizes into a pomeron flux $f(\xp,t)$ and pomeron parton distributions
$q_\funp(\beta,Q^2)$ obeying the DGLAP equations
\be
\label{eq:3}
F^D_2=f(\xp,t)\,\sum_{i=1}^{N_f}e_i^2\,\beta\,\left\{2\, q_\funp(\beta,Q^2)
\right\}\,,
\ee
where $q_\funp=\overline{q}_\funp$ reflects the vacuum nature of the pomeron.
In this model $\beta$ is a fraction of the pomeron momentum carried by a quark.
The QCD analysis of the early HERA data using the IS  model was done in \cite{GK} with the pomeron flux
$
f(\xp,t)\sim \xp^{1-2\alpha_\funp(t)},
$
where the parameters of the Regge trajectory and initial parton distributions
were determined from analyses of
soft hadronic reactions, e.g. the soft pomeron value, $\alpha_\funp(0)=1.1$, was used.
More recent  analyses of inclusive DIS diffraction
\cite{H197,ZEUS99, ROYON} assume that the DPD exhibit the same
factorization as in the IS model, called {\it Regge factorization},
\be
\label{eq:4}
q^D(\xp,t;\beta,Q^2)= {f}(\xp,t)\,\tilde{q}(\beta,Q^2)\,,
\ee
and  parameters in (\ref{eq:4}) (including $\alpha_\funp(0)$) are determined
from  fits to the diffractive DIS data using the DGLAP evolution equations.

In all cases a good description of data is found. However, the basic expe\-ri\-men\-tal facts:
the constant ratio $\sigma^{diff}/\sigma^{tot}$ as a function of energy
and Regge factorization, are described but not understood.

\section{Dipole models and saturation}

\begin{figure}[t]
  \vspace*{-1.0cm}
     \centerline{
         \epsfig{figure=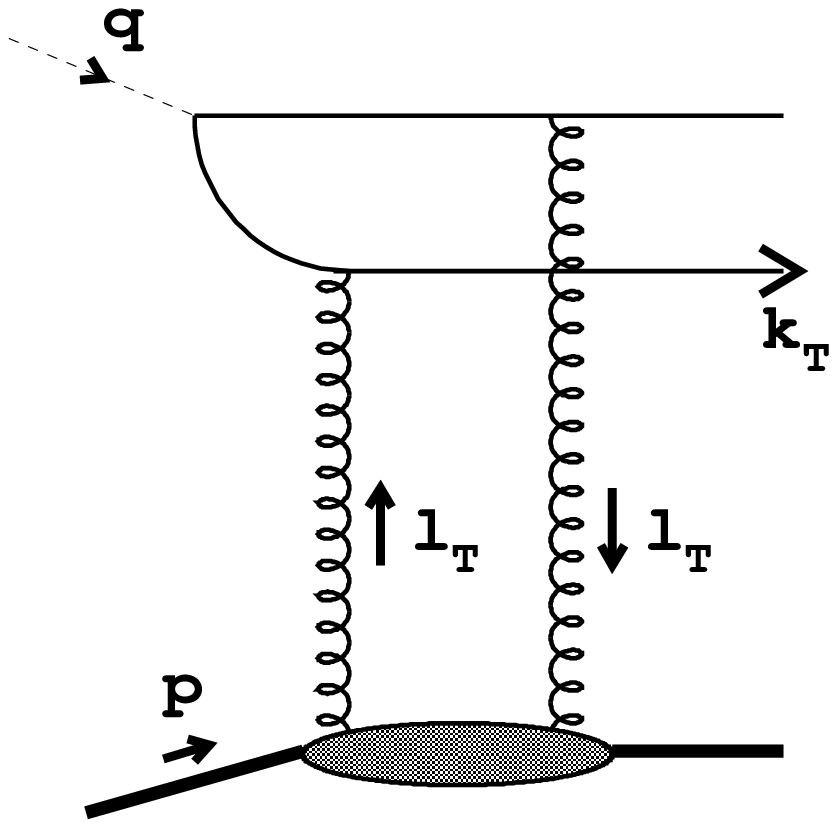,width=5cm}
         \epsfig{figure=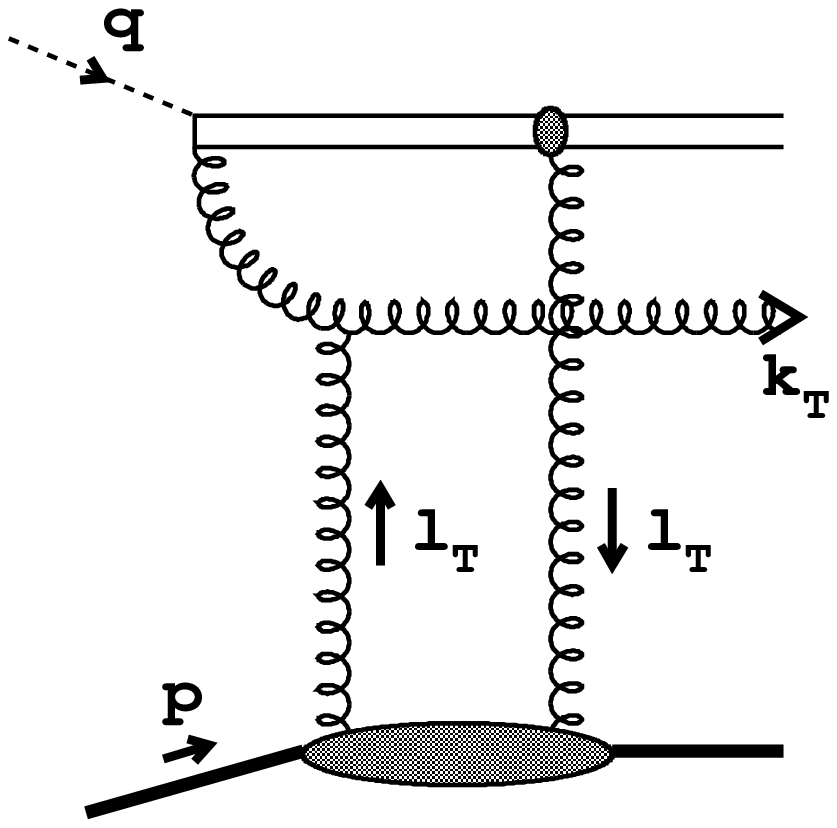,width=5cm}
           }
\vspace*{-0.5cm}
\caption{\it The $q\bar{q}$ and $q\bar{q}g$ components of the diffractive system.
\label{fig:diag1}}
\end{figure}

In these  models, see \cite{BEKW,REVWU,REVHEB}, the diffractive final state is built starting from
a $q\bar{q}$ pair in the color singlet state, and subsequently higher Fock components
($q\bar{q}g$ being the first one)   are added (Fig.~\ref{fig:diag1}).
The colorless interaction of such a diffractive
state with the proton is also modelled. This could be two gluons in the color singlet state
(which leads to no energy dependence) or more complicated gluon exchanges, e.g. the BFKL ladder
with much stronger than the soft pomeron energy dependence. In the simplest case of the $q\bar{q}$
system, the interaction is encoded in a dipole cross section $\hat{\sigma}(x,r)$. The diffractive
$\gamma^*p\to q\bar{q}p^\prime$ cross section is given in this case by
\be
\label{eq:5}
\frac{d\,\sigma^{diff}}{dt}_{\mid\, t=0}
\,=\,
\frac{1}{16\,\pi}\,
\int d^2 r\, dz\,
|\Psi^\gamma(r,z,Q^2)|^2\ \hat\sigma^2(x,r),
\ee
where  $r$ is the transverse
separation of the $q\bar{q}$ pair (dipole),  $z$ is the photon longitudinal momentum fraction
carried by a quark and $\Psi^\gamma$ is the light-cone wave function of the virtual photon.

In \cite{GBW1}  the following form of the dipole cross section was proposed
\be
\label{eq:6}
\hat{\sigma}(x,r)=\sigma_0 \{1-\exp(-r^2 Q^2_s(x))\}\,,
\ee
where the {\it saturation scale} $Q^2_s(x)\sim x^{-\lambda}$ with $\lambda\approx 0.3$.
Three parameters of (\ref{eq:7}) were found from a fit to DIS data on $F_2$
since the same dipole cross section is involved in the description of
$\sigma^{tot}\sim F_2/Q^2$.
Formula (\ref{eq:7}) captures essential features of parton saturation, see \cite{SATREV} and references therein.
In particular, it is important that $\hat{\sigma}\approx \sigma_0$
for $r\gg 1/Q_s(x)$, and that the boundary of this region $1/Q_s(x)\to 0$  for decreasing $x$.
Thus with increasing energy, the dipole cross section saturates (proton is black) for smaller  dipole
sizes.
In the dual momentum space, the dipole cross section
corresponds to the number of gluons per unit of
rapidity and transverse momentum.
With the form (\ref{eq:6}), the number of gluons with
transverse momenta $k_T\gg Q_s(x)$ is proportional to  $1/k_T^2$ and gluons are dilute,
while for small momenta, $k_T\le Q_s(x)$, the number of gluons is tamed by their fusion in a
dense system \cite{GSM}. In such a case,
in the dipole space, $\hat{\sigma}\approx \sigma_0$.
With decreasing $x$, this effect occurs for transverse momenta $k_T\gg \Lambda_{QCD}$.
This is the region were nonlinear QCD evolution equations appear \cite{GLR}.

The DIS diffraction is an ideal process to study parton saturation  since it is
especially sensitive to the large dipole contribution, $r>1/Q_s(x)$.
Unlike inclusive DIS,
the region below is suppressed by an additional power of $1/Q^2$. Moreover, saturation leads in a natural way
to the constant ratio \cite{GBW2}
\be
\label{eq:7}
\frac{\sigma^{diff}}{\sigma^{tot}} \sim \frac{1}{\ln(Q^2/Q^2_s(x))}\,.
\ee
A good
description of diffractive DIS was obtained in this approach without additionally fitted parameters
\cite{GBW2}. For other parameterization
of $\hat{\sigma}$  which describes diffractive data but does not use
the saturation form, see \cite{JEFF}.

The description which is based on the high energy formula (\ref{eq:5}) contains all powers of $1/Q^2$
(twists). Extracting the leading twist contribution from both $q\bar{q}$ and $q\bar{q}g$ components,
the quark and gluon
DPD can  directly be computed  in the saturation model \cite{GBW3}.
An exciting aspect of this calculation is the Regge factorization of the DPD,
\be
\label{eq:8}
\xp\, q^D(\xp,\beta)= Q^2_s(\xp)\,\bar{q}(\beta)\sim \xp^{-0.3}\,,
\ee
due to the form (\ref{eq:6}) in  which  $r$ and $x$  (or $\xp$) are combined
into one dimensionless variable $r Q_s(x)$. This also leads to
the geometric scaling for inclusive DIS \cite{GSCAL}. The
dependence: $F^D_2\sim \xp^{1-2\alpha_\funp}$ with $\alpha_\funp\approx 1.15$, resulting from (\ref{eq:8}),
is in remarkable agreement  with the data \cite{H197,ZEUS99}. Thus the Regge type behaviour and the
dependence on energy of the diffractive DIS data are naturally explained.

\section{Diffractive vector meson production}

\begin{figure}[t]
  \vspace*{0.0cm}
     \centerline{
         \epsfig{figure=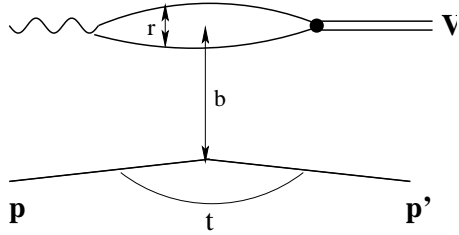,width=6cm}
           }
\vspace*{0.3cm}
\caption{\it Diffractive vector meson production.
\label{fig:vecmes}}
\end{figure}

Diffractive vector meson production gives an access to more detailed structure of the $q\bar{q}$ dipole
interaction with the proton. Namely, the dipole--proton scattering amplitude $N(r,b,x)$ can be studied,
for which
\be
\label{eq:9}
\hat{\sigma}(r,x)=2 \int d^2 b\,N(r,b,x)\,,
\ee
where $b$ is the impact parameter of the dipole, see Fig.~\ref{fig:vecmes}.
Through the $t$-dependence (at small $t$) of the vector meson production cross section,
the impact parameter dependence of this amplitude can be analyzed since
\be
\label{eq:10}
\frac{d\sigma^{VM}}{dt}=\frac{1}{16\pi}
\left|\,\Psi^V\, \otimes \int d^2 b\,\mbox{\rm e}^{ib \cdot \Delta}\, N(r,b,x)
\,\otimes\, \Psi^\gamma\, \right|^2
\ee
where $\Delta$ is a two-dimensional vector of transverse momentum transferred into a vector meson:
$t=-\Delta^2$. Formula (\ref{eq:10}) reflects the three step factorization, shown in Fig.~\ref{fig:vecmes},
and involves  a nonperturbative vector meson wave function  $\Psi^V$, which needs to be modelled.
The first studies of the diffractive $J/\psi$ production in the presented approach
has already been performed \cite{KOWALSKI}.

The amplitude $N$ can also be obtained from the QCD nonlinear evolution equation
of Balitsky and Kovchegov \cite{BAL}, resulting from Color Glass Condensate,
an effective theory of dense gluon systems with saturation \cite{SATREV}. This is an exciting
program to confront the theoretical studies of saturation using the BK equation with the
phenomenological analysis of data.

\medskip
\centerline{ACKNOWLEDGEMENTS}

I dedicate this presentation to the memory of Professor Jan Kwieci\'nski, my teacher and master.
A partial support by the KBN grant No. 5 P03B 144 20 is acknowledged.


\end{document}